# Response to "Verifying quantum superpositions at metre scales"

T. Kovachy, P. Asenbaum, C. Overstreet, C. A. Donnelly, S. M. Dickerson, A. Sugarbaker, J. M. Hogan, and M. A. Kasevich

*Department of Physics, Stanford University, Stanford, California 94305*

**The preceding BCA[1] asserts that our observation of interference contrast in a half-metre-scale atom interferometer[2] does not prove the existence of macroscopic quantum superpositions and hence does not test quantum mechanics at long length scales. Moreover, the BCA implies that intrinsic atomic interactions or technical imperfections could prevent the application of our work to future differential measurements. On the contrary, we argue the following: i) in standard quantum mechanics, there is no known mechanism in our system that prohibits its use in future differential measurement applications; ii) our experiment tests quantum mechanics in that it constrains any modifications that would reduce contrast in an interferometer with arms that propagate over widely separated trajectories; and iii) using a standard definition of superposition, our observation of interference results from quantum superposition at the half-metre scale. In particular, quantum superposition is a more general concept than first-order coherence.**

We operated our atom source with a condensate fraction of $\sim 50\%$. The atomic source has a coherence length of only $2 \times 10^{-6}$ m, substantially smaller than the spatial extent of the cloud. This arises from imperfections in the magnetic lensing, lattice launch, and atom optics interactions. Coherence between the two interferometer arms is established by the initial beam splitter pulse, at which time the ratio U/J of the interaction matrix element to Bragg transition Rabi frequency is $\sim 10^{-8}$, ruling out interaction-based effects during the beam splitter[3]. The atomic density is no larger than $\sim 10^{10}/\text{cm}^3$ during the interferometer sequence, which is dilute enough to prevent dephasing due to mean-field shifts (mean-field shift $\sim 0.1$ Hz). Under these conditions, standard quantum mechanics rules out evolution into the state described in the BCA. Furthermore, we know of no technical noise sources that will lead to emergence of such states (all known technical noise sources, such as residual spontaneous emission, are associated with momentum exchange which modifies the structure of the atomic states and reduces contrast). Thus, there is no known mechanism that would prohibit the utilization of the acceleration sensitivity inferred from the large arm separation in differential measurement applications, such as our current work with dual species interferometry for an equivalence principle test[4].

When evaluating the degree to which our experiment constrains a particular hypothetical modification of quantum mechanics, it is important to consider disturbances to the states of individual atoms – for example, due to momentum exchange that fundamentally alters the structure of the many-atom state as the state propagates. In the case of momentum exchange, the large spatial separation directly translates into an increased sensitivity to this spurious heating. A spurious momentum kick $\hbar q$, if it occurs midway through the interferometer, is associated with a wavepacket phase shift of $\left[ m(v+\hbar q/m)^2/2 - mv^2/2 \right] T/\hbar \sim qL$, where $m$ is the atomic mass, $v$ is the velocity separation, $T$ is the drift time, and $L \sim vT$ is the wavepacket separation. Even momentum kicks as small as $q \sim 2\pi/L$ (corresponding to wavelength $\sim L$) result in phase shifts of $\sim 2\pi$ rad, which, if they occur inhomogeneously, result in reduced contrast. Modifications that only add overall phase noise are not ruled out by our results.

We would like to clarify our use of the word "superposition." In our paper, following Feynman and others, we adopted the nomenclature that interference – whether or not there is a determinate phase – necessarily results from superposition[5,6]. This view of superposition is elegantly illustrated by the Pfleegor-Mandel experiment[7], which tracks the buildup of an interference pattern from two independent laser beams one photon at a time. A standard interpretation of these experiments is that interference results from superposition between the sources, which is revealed during the detection process[7].